\documentclass[9pt,conference]{IEEEtran}
\usepackage{waspaa25} % removes hyperlinks (required by IEEE Xplore)

\usepackage{bm} % for bold math symbols (incl. Greek letters) with \bm{}
\usepackage{mathrsfs}
\usepackage{bbm}
\usepackage{amsxtra}
\usepackage{threeparttable}
\usepackage{algorithm,algorithmic}
\usepackage{multirow}
\usepackage{textcomp}
\usepackage{xcolor}
\usepackage{pifont}
\usepackage[inkscapelatex=false]{svg}
\title{Physics-Informed Direction-Aware Neural Acoustic Fields}

\name{Yoshiki Masuyama$^{1}$,
      François G.\ Germain$^{1}$,
      Gordon Wichern$^{1}$,
      Christopher Ick$^{1, 2}$,
      Jonathan Le Roux$^{1}$}
\address{$^{1}$Mitsubishi Electric Research Laboratories (MERL), Cambridge, MA, USA \; \\
$^{2}$Music and Audio Research Laboratory, New York University, NY, USA
}

\begin{document}
\maketitle

\begin{abstract}
This paper presents a physics-informed neural network (PINN) for modeling first-order Ambisonic (FOA) room impulse responses (RIRs).
PINNs have demonstrated promising performance in sound field interpolation by combining the powerful modeling capability of neural networks and the physical principles of sound propagation.
In room acoustics, PINNs have typically been trained to represent the sound pressure measured by omnidirectional microphones where the wave equation or its frequency-domain counterpart, i.e., the Helmholtz equation, is leveraged.
Meanwhile, FOA RIRs additionally provide spatial characteristics and are useful for immersive audio generation with a wide range of applications.
In this paper, we extend the PINN framework to model FOA RIRs.
We derive two physics-informed priors for FOA RIRs based on the correspondence between the particle velocity and the $(X, Y, Z)$-channels of FOA.
These priors associate the predicted $W$-channel and other channels through their partial derivatives and impose the physically feasible relationship on the four channels.
Our experiments confirm the effectiveness of the proposed method compared with a neural network without the physics-informed prior.
\end{abstract}

%%%%%%%%%%%%%%%%%%%%%%
\section{Introduction}
\label{sec:intro}
%%%%%%%%%%%%%%%%%%%%%%

Room impulse responses (RIRs) capture sound propagation in indoor environments and serve as a fundamental component of room acoustics~\cite{kuttruff2016}.
In addition to the direct sound, an RIR involves reflections and diffraction depending on the materials of the room surface and the positions of the sound source and microphone.
RIRs dramatically vary in each room, and their precise modeling is crucial for high-quality immersive audio experiences in mixed reality applications~\cite{lentz2007,serafin2018}.
Furthermore, RIR modeling is useful to enhance various downstream tasks, including dereverberation~\cite{bahrman24}.
RIRs are typically measured by using a loudspeaker and one or more microphones placed within a room.
The measurement process requires an enormous amount of time and effort to cover a large area with dense spatial grids.
To alleviate this issue, spatial interpolation from a set of sparse measurements is in great demand, and various techniques have been developed~\cite{ueno2018,lee2017,tsunokuni2021,jalmby2023,sundstrom2024}, notably using compressed sensing~\cite{mignot2013,antonello2017,verburg2018}.

Recently, neural networks have been leveraged in RIR interpolation and generation, motivated by their adaptive and powerful modeling capabilities~\cite{bryan2020,lluis2020,ratnarajah2021,pezzoli2022,ratnarajah2022,chen2024RAF}.
In particular, neural fields (NFs) have gained increasing attention due to their flexibility~\cite{luo22naf,richard2022,su22inras,liang23avnerf}.
An NF is trained to predict the sound pressure from a given microphone position and optionally other contexts, e.g., the source position, on the sparse measurements.
Once it is trained, we can predict RIRs at arbitrary positions in a grid-less manner.
However, NFs are typically trained from scratch for each room, which leads to overfitting and results in poor generalization capability to unmeasured positions.
More fundamentally, NFs do not respect the physical principles of sound propagation when trained solely to minimize reconstruction error.

Physics-informed neural networks (PINNs)~\cite{raissi2019,karniadakis2021} have mitigated this issue by enforcing the network output to follow the physics of sound propagation~\cite{olivieri2021,chen2023,pezzoli2023,Karakonstantis2024,sato2024,koyama2024pinnsound}.
By exploiting automatic differentiation, PINNs efficiently calculate the derivatives of the network outputs with respect to the inputs, e.g., the microphone position.
During training, the NF is then regularized to minimize the discrepancy of the network outputs and their derivatives from the governing partial differential equation (PDE).
For the sound field reconstruction, the time-domain wave equation or its frequency-domain counterpart, i.e., the Helmholtz equation, have been widely used.

Along with the modeling of omnidirectional RIRs, which is now a mature field, the modeling of spatial RIRs has gained increasing attention to render binaural audio according to listener orientation~\cite{merimaa2005,coleman2017,Zaunschirm2018}.
In particular, first-order Ambisonics (FOA)~\cite{gerzon1973,daniel1998,arteaga2023} encodes the directional characteristics of sound field based on spherical harmonic decomposition~\cite{williams1999}.
The zeroth-order component known as the $W$-channel corresponds to the sound pressure measured by an omnidirectional microphone.
Meanwhile, the first-order components, i.e., the $(X, Y, Z)$-channels, encode the particle velocity along the Cartesian directions.
The generation and interpolation of FOA RIRs have improved with the use of neural networks~\cite{ren2024,danf}.
In particular, the direction-aware neural acoustic field (DANF) learns to predict the FOA RIR, i.e., the $(W, X, Y, Z)$-channels, from the microphone position and the context, facilitating grid-less reconstruction of spatial RIRs~\cite{danf}.
Consequently, DANF can be used to efficiently decode binaural audio depending on listener orientation at any position.
DANF optimization has relied solely on data fidelity to sparsely measured FOA RIRs, and it should be beneficial to consider physical principles.

\begin{figure}[t!]
\centering
 \includegraphics[width=0.99\linewidth]{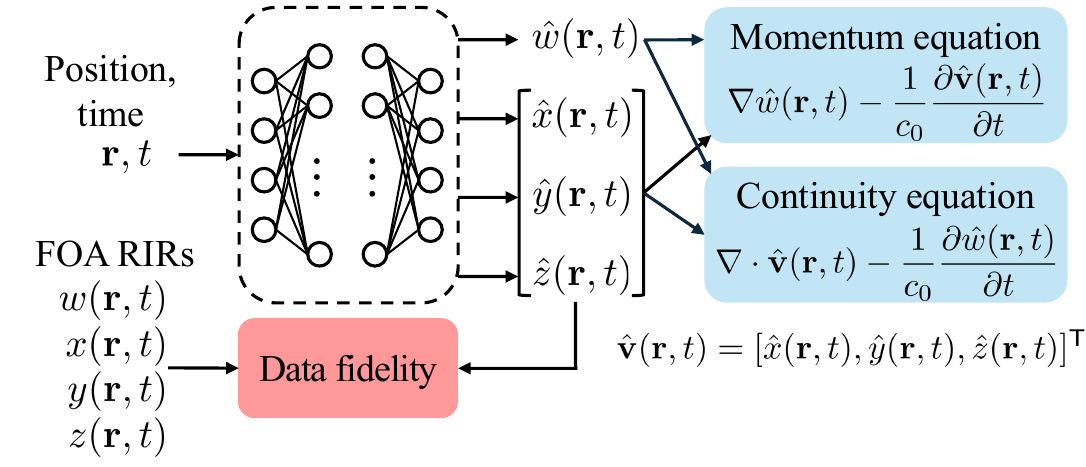}
\vspace{-10pt}
\caption{Illustration of the proposed PINN for FOA RIR modeling, PI-DANF.
The blue boxes show the physics-informed priors tailored for Ambisonics.}
\label{fig:overview}
\end{figure}

In this paper, we propose PI-DANF, a physics-informed extension of DANF to generate acoustically valid FOA RIRs.
We leverage the analytical relationship between the $(X, Y, Z)$-channels and the particle velocity, and derive physics-informed priors from the linearized momentum equation and the continuity equation as illustrated in \cref{fig:overview}.
Since the wave equation is derived by combining these PDEs, PI-DANF can be interpreted as a natural extension of the PINN used for omnidirectional RIR~\cite{Karakonstantis2024}.
Our experiments on simulated data validate the effectiveness of PI-DANF compared to the vanilla DANF and to DANF with the previous wave-equation-based prior on the $W$-channel.

%%%%%%%%%%%%%%%%%%%%%%%%%
\section{Preliminaries}
\label{sec:Preliminaries}
%%%%%%%%%%%%%%%%%%%%%%%%%

%%%%%%%%%%%%%%%%%%%%%%%%%%%%%%%%%%%%%%%%%%%%%%%%%%%%%%%%%%%%%
\subsection{Sound Field Reconstruction Using Neural Networks}
%%%%%%%%%%%%%%%%%%%%%%%%%%%%%%%%%%%%%%%%%%%%%%%%%%%%%%%%%%%%%

Let us denote the sound pressure $p(\mathbf{r}, t) \in \mathbb{R}$ as a function of position $\mathbf{r} \in \mathbb{R}^3$ in Cartesian coordinates and time $t \in \mathbb{R}$.
In particular, we consider the sound field in a region $\Omega \subset \mathbb{R}^3$ that does not contain sources.
The sound pressure is assumed to follow two PDEs under regular conditions~\cite{blackstock2000}.
The first equation is the linearized momentum equation given by 
\begin{equation}
    \nabla p(\mathbf{r}, t) + \rho_0 \frac{\partial \mathbf{u}(\mathbf{r}, t)}{\partial t} = \mathbf{0}, \label{eq:momentum}
\end{equation}
where $\mathbf{u}(\mathbf{r}, t) \in \mathbb{R}^3$ is the particle velocity, $\nabla$ denotes the gradient with respect to $\mathbf{r}$, and $\rho_0$ is the density of the air.
The other equation is derived from the continuity equation as follows:
\begin{equation}
\rho_0 \nabla \cdot \mathbf{u}(\mathbf{r}, t) + \frac{1}{c_0^2} \frac{\partial p(\mathbf{r}, t)}{\partial t} = 0, \label{eq:continuity}
\end{equation}
where $\nabla \cdot$ denotes the divergence, and $c_0$ is the sound speed.
On the basis of \cref{eq:momentum}--\cref{eq:continuity}, the wave equation is derived as follows:
\begin{equation}
\Delta p(\mathbf{r}, t) - \frac{1}{c_0^2} \frac{\partial^2 p(\mathbf{r}, t)}{\partial t^2}= 0,
\label{eq:waveeq}
\end{equation}
where $\Delta$ denotes the Laplacian.

We aim to reconstruct the continuous sound field $p(\mathbf{r}, t)$ from its sparse measurements at $\mathbf{r}_d \in \Omega$, where $d = 0, \ldots, D-1$ is the index of the microphone position.
Sound field reconstruction can be realized by optimizing an NF to map the given position and time to the corresponding sound pressure~\cite{richard2022,su22inras}:
\begin{equation}
    p(\mathbf{r}, t) \approx \hat{p}(\mathbf{r}, t) = \texttt{NF}_\theta(\mathbf{r}, t),
    \label{eq:mononf}
\end{equation}
where $\theta$ denotes the parameters of the NF.
The NF is trained to minimize the following reconstruction error of the measurements:
\begin{equation}
     \mathcal{L}_\text{data} = \frac{1}{DL}\sum_{d=0}^{D-1} \sum_{l = 0}^{L-1} |\hat{p}(\mathbf{r}_d, t_l) - p(\mathbf{r}_d, t_l)|,
     \label{eq:datafidelity}
\end{equation}
where $l = 0, \ldots, L-1$ is the sample index of the measured RIRs.
Once the NF is trained, it can predict sound pressure at any position and time in a grid-less manner.

The training of NFs solely with the loss function in \cref{eq:datafidelity} does not consider the properties of sound propagation, resulting in poor generalization capability to unmeasured positions.
The PINN framework mitigates this issue by incorporating a prior term derived from the physical principles of the sound propagation.
Specifically, the prior term derived from the wave equation in \cref{eq:waveeq} has been widely used for time-domain modeling~\cite{pezzoli2023,Karakonstantis2024}:
\begin{equation}
     \mathcal{L}_\text{wave} =  \mathbb{E}_{\mathbf{r} \in \Omega} \mathbb{E}_{t \in [0, T]} \left| \Delta \hat{p}(\mathbf{r}, t) - \frac{1}{c_0^2} \frac{\partial^2 \hat{p}(\mathbf{r}, t)}{\partial t^2}  \right|,
     \label{eq:waveeqloss}
\end{equation}
where $\mathcal{L}_\text{wave}$ measures the discrepancy from the wave equation.
This prior term can be efficiently computed at any position $\mathbf{r} \in \Omega$ and time $t \in [0, T]$ by using automatic differentiation, where $T \in \mathbb{R}_+$ is for limiting the time range.
PINNs have demonstrated promising performance under various conditions compared with compressed sensing and other deep learning methods~\cite{pezzoli2023}.

%%%%%%%%%%%%%%%%%%%%%%%%%%%%%%%%%%%%%%%%%
\subsection{First-Order Ambisonics (FOA)}
\label{sec:foa}
%%%%%%%%%%%%%%%%%%%%%%%%%%%%%%%%%%%%%%%%%

FOA, often referred to as simply Ambisonics, is an audio format that captures the spatial characteristics of a sound field by using the spherical harmonics up to the first order~\cite{gerzon1973,daniel1998,arteaga2023}.
This representation is beneficial for a wide range of applications, including mixed reality, and has been used for 3D spatial audio coding~\cite{herre2014}.
FOA encodes the sound field into four channels, where the $W$-channel matches the sound pressure measured by a virtual omnidirectional microphone up to a multiplicative constant.
Meanwhile, the $(X, Y, Z)$-channels are derived from the spherical harmonics of the first order and can be interpreted as measurements from virtual dipole microphones along the Cartesian directions.
Under the SN3D normalization~\cite{daniel2003}, the sound pressure $p(\mathbf{r}, t)$ is equivalent to $w(\mathbf{r}, t)$, and the particle velocity $\mathbf{u}(\mathbf{r}, t)$ is given by
\begin{align}
\mathbf{u}(\mathbf{r}, t) &= -\frac{1}{ \rho_0 c_0} \mathbf{v}, \label{eq:velocity} \\
\mathbf{v}(\mathbf{r}, t) &= [x(\mathbf{r}, t), y(\mathbf{r}, t), z(\mathbf{r}, t)]^\mathsf{T},
\end{align}
where $w(\mathbf{r}, t)$, $x(\mathbf{r}, t)$, $y(\mathbf{r}, t)$, and $z(\mathbf{r}, t)$ are the channels of the FOA RIR measured at position $\mathbf{r}$ and time $t$~\cite{merimaa2005}.

Recently, deep generative models have been developed for audio in the FOA format~\cite{heydari2025,kushwaha2025,kim2025}. Related to our work, a paper proposes a generative adversarial network for FOA RIRs~\cite{ren2024}.
These previous studies mainly focus on generating plausible audio, while we aim to interpolate FOA RIRs in a physically feasible manner.

%%%%%%%%%%%%%%%%%%%%%%%%%
\section{Proposed Method}
%%%%%%%%%%%%%%%%%%%%%%%%%

%%%%%%%%%%%%%%%%%%%%%%%%%%%%%%%%%%%%%%%%%%%%%%%%
\subsection{Direction-Aware Neural Field (DANF)}
%%%%%%%%%%%%%%%%%%%%%%%%%%%%%%%%%%%%%%%%%%%%%%%%

While the $(X, Y, Z)$-channels of FOA RIRs can be computed using a PINN for sound pressure in \cref{eq:mononf} through the linearized momentum equation in \eqref{eq:momentum}, this requires the numerical integration of the spatial gradient over time, which makes binaural decoding too complex~\cite{Karakonstantis2024}.
As an alternative, DANF has been proposed to extend an NF from predicting omnidirectional RIRs to representing FOA RIRs~\cite{danf}:
\begin{equation}
    \hat{w}(\mathbf{r}, t), \hat{\mathbf{v}}(\mathbf{r}, t) = \texttt{DANF}_\theta(\mathbf{r}, t).
\end{equation}
The network is optimized by minimizing the reconstruction error of sparse measurements%
\footnote{The original paper also employs additional losses in the short-time Fourier transform domain, but all the losses concern the fidelity to the measured FOA RIRs~\cite{danf}.}:
\begin{align}
     \mathcal{L}_\text{data} &= \frac{1}{DL}\sum_{d=0}^{D-1} \sum_{l = 0}^{L-1}  ( |\hat{w}(\mathbf{r}_d, t_l) - w(\mathbf{r}_d, t_l)| \nonumber \\
     &\hspace{102pt} + \|\hat{\mathbf{v}}(\mathbf{r}_d, t_l) - \mathbf{v}(\mathbf{r}_d, t_l)\|_1 ),
     \label{eq:datafidelity-foa}
\end{align}
where $\| \cdot \|_1$ denotes the $\ell_1$ norm.
DANF directly provides spatial information at the given position, which allows binaural decoding depending on listener orientation at any position.

In the original paper~\cite{danf}, DANF struggled to predict reliable FOA RIRs at unmeasured positions when trained from scratch with a limited number of measurements.
Although pre-training DANF on large-scale RIR datasets was proposed to alleviate this issue, it does not guarantee adherence to the physical principles of sound propagation.
We expect the PINN framework to address this problem.

%%%%%%%%%%%%%%%%%%%%%%%%%%%%%%%%%%%%%%%%%%%%%%%%%%
\subsection{Physics-Informed Prior for Ambisonics}
%%%%%%%%%%%%%%%%%%%%%%%%%%%%%%%%%%%%%%%%%%%%%%%%%%

A naive way to adopt a PINN framework for DANF is to apply the wave-equation-based prior in \cref{eq:waveeqloss} to the $W$-channel because that channel corresponds to the sound pressure.
This training, however, only regularizes the $W$-channel and leaves the $(X, Y, Z)$-channels unconstrained.
Thus, it may not be sufficient to interpolate FOA RIRs with appropriate spatial characteristics.

We thus propose two physics-informed priors tailored for FOA RIRs.
As described in \cref{sec:foa}, the $(X, Y, Z)$-channels are analytically related to the particle velocity in \cref{eq:velocity}.
By substituting $\hat{w}(\mathbf{r}, t)$ and $-\hat{\mathbf{v}}(\mathbf{r}, t)/(c_0 \rho_0)$ respectively into $p(\mathbf{r}, t)$ and $\mathbf{u}(\mathbf{r}, t)$ in the linearized momentum equation \cref{eq:momentum}, the first penalty term is formulated as
\begin{equation}
\mathcal{L}_\text{momentum} = \mathbb{E}_{\mathbf{r} \in \Omega} \mathbb{E}_{t \in [0, T]} \left\| \nabla \hat{w}(\mathbf{r}, t) - \frac{1}{c_0} \frac{\partial \hat{\mathbf{v}}(\mathbf{r}, t)}{\partial t} \right\|_1.
\label{eq:momentumloss}
\end{equation}
The other prior is derived by substituting the network outputs into the continuity equation \cref{eq:continuity} and multiplying by $c_0$:
\begin{equation}
    \mathcal{L}_\text{continuity} = \mathbb{E}_{\mathbf{r} \in \Omega} \mathbb{E}_{t \in [0, T]} \left|
    \nabla \cdot \hat{\mathbf{v}}(\mathbf{r}, t) - \frac{1}{c_0} \frac{\partial \hat{w}(\mathbf{r}, t)}{\partial t}
    \right|.
    \label{eq:continuityloss}
\end{equation}
These proposed priors connect the predictions of all channels, in contrast to the wave-equation-based prior which focuses solely on the $W$-channel.
Furthermore, they penalize the absolute value of four scalars per position $\mathbf{r}$ and time $t$, while the previous one penalizes one scalar.
That is, the proposed priors more precisely reflect the physical principles of FOA RIRs.

%%%%%%%%%%%%%%%%%%%%%%%%%%%%%%%%%%%%%%%%%%%%%%%%%%%%
\subsection{Network Architecture and Training Setup}
%%%%%%%%%%%%%%%%%%%%%%%%%%%%%%%%%%%%%%%%%%%%%%%%%%%%

To realize PI-DANF, we follow the network architecture and training scheme of the PINN for sound pressure~\cite{Karakonstantis2024}.
For the network, the modified multi-layer perceptron is used to perform adaptive processing.
The network input $\mathbf{h}_0 = [\mathbf{r}^\mathsf{T}, t]^\mathsf{T}$ is passed to two encoders:
\begin{align}
    \mathbf{f} &= \texttt{SIREN}_f(\mathbf{h}_0), \\
    \mathbf{g} &= \texttt{SIREN}_g(\mathbf{h}_0),
\end{align}
where $\texttt{SIREN}$ is the fully connected layer with sine activation~\cite{Sitzmann2020}.
This is advantageous for PINN because the periodic activation function provides well-behaved derivatives.
Then, the forward process of the $k$-th hidden layer is defined as 
\begin{align}
    \tilde{\mathbf{h}}_k &= \texttt{SIREN}_k(\mathbf{h}_{k-1}), \\
    \mathbf{h}_k &= (1 - \tilde{\mathbf{h}}_k) \odot \mathbf{f} + \tilde{\mathbf{h}}_k \odot \mathbf{g},
\end{align}
where $\odot$ denotes Hadamard product, and $k = 1, \ldots, K$.
The output of the final hidden layer $\mathbf{h}_K$ is then projected to obtain the FOA RIR estimate $[\hat{w}(\mathbf{r}, t), \hat{\mathbf{v}}(\mathbf{r}, t)^\mathsf{T}]^\mathsf{T}$.

The network is then optimized to minimize the data fidelity term and the sum of the proposed priors with variable weights $(\epsilon_\text{data}, \epsilon_\text{prior})$:
\begin{equation}
    \!\! \mathcal{L}_\text{all} = \frac{1}{2 \epsilon_\text{data}^2} \mathcal{L}_\text{data} + \frac{1}{2 \epsilon_\text{prior}^2} (\mathcal{L}_\text{momentum} + \mathcal{L}_\text{continuity}) + \log (\epsilon_\text{data} \epsilon_\text{prior}), \!\!
    \label{eq:adaptive}
\end{equation}
where the weights are adaptively updated together with the network parameters $\theta$~\cite{xiang2022}.
The limit of $\epsilon_\text{prior} \rightarrow \infty$ corresponds to the vanilla DANF without the physics-informed priors.
For the priors, we stochastically take the position $\mathbf{r} \in \Omega$ and time $t \in [0, T]$ in each iteration by using Latin hypercube sampling as in \cite{Karakonstantis2024}.

%%%%%%%%%%%%%%%%%%%%%
\section{Experiments}
%%%%%%%%%%%%%%%%%%%%%

%%%%%%%%%%%%%%%%%%%%%%%%%%%%%%%%%%%%%%%%%%%
\subsection{Dataset and experimental setup}
%%%%%%%%%%%%%%%%%%%%%%%%%%%%%%%%%%%%%%%%%%%

\begin{figure}[t!]
\centering
 \includegraphics[width=0.9\linewidth]{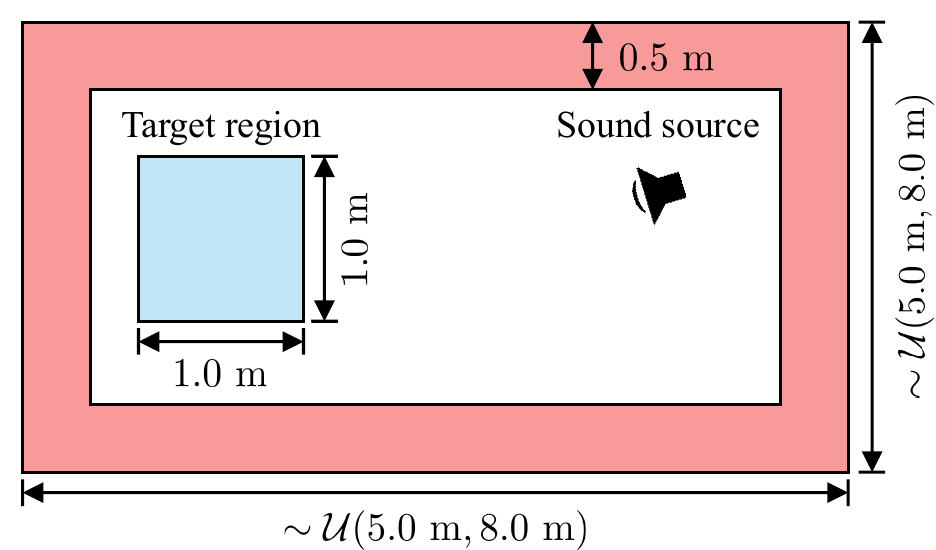}
\caption{Room geometry for simulation.
The room height was sampled from $\mathcal{U}(2.5~\text{m}, 4.0~\text{m})$.}
\label{fig:exproom}
\end{figure}

We evaluate the proposed PI-DANF with FOA RIRs simulated by \texttt{HARP}%
\footnote{\url{https://github.com/whojavumusic/HARP}}
~\cite{saini2024}.
\texttt{HARP} is built on top of \texttt{Pyroomacoustics}~\cite{Scheibler2018PyRoom} and generates high-order Ambisonics RIRs by combining the image source method with specific directivity patterns.
While the original dataset provides $100,000$ RIRs under various room conditions, we simulated FOA RIRs for $10$ rooms with a random geometric setup and used their early part up to $100$ ms with a sampling rate of $8$ kHz.
Each room was a shoebox as depicted in \cref{fig:exproom}, and the target region $\Omega$ was a cuboid of dimensions $1.0$ m $\times$ $1.0$ m $\times$ $1.0$ m.
The sound source was randomly placed anywhere in the room except for the red area near the walls.
We simulated impulse responses on a grid of $5$ cm width in Cartesian coordinates, resulting in $21 \times 21 \times 21 = 9,261$ measurements.
We randomly chose $\{250, 500\}$ measurements to train the NFs and used $50$ other measurements for validation.
We evaluated the network on the remaining $8,711$ measurements.

Our network architecture and training setups mainly followed an implementation of the PINN for omnidirectional RIRs%
\footnote{\url{https://github.com/xefonon/RIRPINN/tree/main}}~\cite{Karakonstantis2024}.
The NFs consisted of 3 hidden layers, i.e., $K=3$, with $512$ units.
They were optimized with the Adam optimizer and cosine annealing.
In each iteration, we used all the measured positions $(\mathbf{r}_d)_{d=0}^{D-1}$ but randomly sampled $250$ discrete times $t_l$ to compute the data fidelity term.
Meanwhile, for the physics-informed priors, $25,000$ pairs of $\mathbf{r}$ and $t$ were taken by Latin hypercube sampling.
We set the initial values of $\epsilon_\text{data}$ and $\epsilon_\text{prior}$ to $1.0$ and $0.1$, respectively.
The NFs were trained for $100,000$ iterations.

The predicted FOA RIRs were evaluated by the normalized mean squared error (NMSE) for the $W$-channel and $(X, Y, Z)$-channels:
\begin{equation}
    \text{NMSE}(w, \hat{w}) = 10 \log_{10} \left(
    \frac{\sum_{\tilde{d}=0}^{\tilde{D}-1} \|\hat{\mathbf{w}}(\mathbf{r}_{\tilde{d}}) - \mathbf{w}(\mathbf{r}_{\tilde{d}})\|_2^2}{\sum_{\tilde{d}=0}^{\tilde{D}-1} \|\mathbf{w}(\mathbf{r}_{\tilde{d}})\|_2^2}\right),
\end{equation}
where $\tilde{d} = 0, \ldots, \tilde{D}$ indexes the evaluation positions, $\mathbf{w}(\mathbf{r}_{\tilde{d}}) = [w(\mathbf{r}_{\tilde{d}}, t_0), \ldots, w(\mathbf{r}_{\tilde{d}}, t_{L-1})]^\mathsf{T}$, and $\|\cdot\|_2$ denotes the $\ell_2$ norm.
A recent study indicates that the correlation to the reference RIR at the $W$- and $(X, Y, Z)$-channels match the listening quality and perceptual localization accuracy~\cite{ren2024icassp}.
We thus measured the Pearson’s correlation coefficients as follows:
\begin{equation}
    \!\!\text{PCC}(w, \hat{w})
    = \frac{1}{\tilde{D}} \sum_{\tilde{d}=0}^{\tilde{D}-1} 
     \frac{[\hat{\mathbf{w}}(\mathbf{r}_{\tilde{d}}) - \hat{\underline{\mathbf{w}}}(\mathbf{r}_{\tilde{d}})]^\mathsf{T}
     [{\mathbf{w}}(\mathbf{r}_{\tilde{d}}) - \underline{\mathbf{w}}(\mathbf{r}_{\tilde{d}})]}{
     \|\hat{\mathbf{w}}(\mathbf{r}_{\tilde{d}}) - \hat{\underline{\mathbf{w}}}(\mathbf{r}_{\tilde{d}})\|_2
     \|{\mathbf{w}}(\mathbf{r}_{\tilde{d}}) - \underline{\mathbf{w}}(\mathbf{r}_{\tilde{d}})\|_2
     },\!\!
\end{equation}
where $\hat{\underline{\mathbf{w}}}(\mathbf{r}_{\tilde{d}})$ and $\underline{\mathbf{w}}(\mathbf{r}_{\tilde{d}})$ are the time average of $\hat{w}(\mathbf{r}_{\tilde{d}}, t_l)$ and $w(\mathbf{r}_{\tilde{d}}, t_l)$, respectively.
The evaluation was performed with the checkpoint that achieved the best NMSE for the $W$-channel on the validation set.

\begin{figure*}
    \centering
    \includegraphics[width=0.99
    \linewidth]{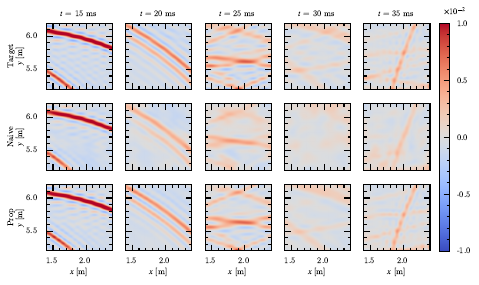}
    \caption{Visualization of the original and predicted sound pressure, i.e., $W$-channel, within the target region at $2.30$ m height.
    The sound source was located at ($0.85$ m, $1.95$ m, $1.31$ m) inside a room of dimensions of ($6.65$ m, $7.15$ m, $3.71$ m).
    }
    \label{fig:soundfield}
\end{figure*}

%%%%%%%%%%%%%%%%%%%%
\subsection{Results}
%%%%%%%%%%%%%%%%%%%%

\Cref{fig:soundfield} compares the sound pressure predicted by the DANF with the wave-equation-based prior on the $W$-channel (Naive) and the proposed PI-DANF (Prop).
These NFs were optimized with $250$ measurements.
According to the leftmost panels, both NFs successfully captured the direct sound and early reflection.
Meanwhile, the naive method resulted in lower power and even missed some reflection after $25$ ms.
The proposed method, on the other hand, achieved better reconstruction.

\begin{figure}[t!]
\centering
\includegraphics[width=0.99\columnwidth]{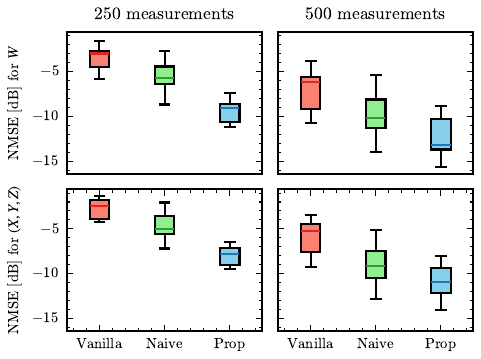}
\vspace{-4pt}
\caption{Boxplots of NMSE over 10 rooms (lower is better).}
\label{fig:nmse}
\end{figure}

\begin{figure}[t!]
\centering
\includegraphics[width=0.99\columnwidth]{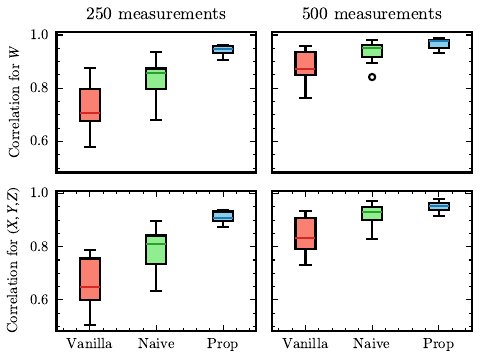}
\vspace{-4pt}
\caption{Boxplots of Pearson’s correlation coefficients (higher is better).}
\label{fig:pcc}
\end{figure}

\Cref{fig:nmse} and \Cref{fig:pcc} depict the NMSE and Pearson’s correlation coefficient for the $W$- and $(X, Y, Z)$-channels.
Here, we averaged the channel-wise scores for the $(X, Y, Z)$-channels.
The naive method outperformed the vanilla DANF that was trained solely with the data fidelity term.
Interestingly, the performance on the $(X, Y, Z)$-channels was also improved even though the wave-equation-based prior regularizes only the $W$-channel.
This could be because the network is shared across different channels except for the final projection layer.
The PI-DANF consistently performed the best owing to the proposed priors tailored for FOA RIRs.

%%%%%%%%%%%%%%%%%%%%
\section{Conclusion}
%%%%%%%%%%%%%%%%%%%%

We proposed PI-DANF, a PINN framework for predicting FOA RIRs.
Based on the linearized momentum equation and the continuity equation, we derived two physics-informed priors to enforce the network outputs to follow the physical principles of FOA RIRs.
These priors connect the $W$-channel and $(X, Y, Z)$-channels through their partial derivatives while the wave-equation-based prior has been applied to only the sound pressure, i.e., $W$-channel.
We have confirmed the effectiveness of the PI-DANF through numerical experiments compared with the vanilla DANF and its naive extension.
Future work will focus on scaling PI-DANF to model complete FOA RIRs, instead of only the early part, in realistic rooms and aim to generate physically feasible and perceptually natural RIRs.

% The \IEEEtriggeratref{XX} command can be used to move to the next column before the XX-th reference
% to balance the two columns of the reference section
% \IEEEtriggeratref{XX}
\bibliographystyle{IEEEtran}
\bibliography{refs25}

\end{document}